\documentclass[a4paper,UKenglish,cleveref, autoref, thm-restate]{lipics-v2021}

\pdfoutput=1 
\hideLIPIcs  

\usepackage[font=footnotesize,labelfont=bf]{caption}
\usepackage[font=scriptsize,labelfont=bf]{subcaption}
\usepackage{mathtools}
\usepackage{xfrac}
\usepackage{bm}
\usepackage{tikz}
\usetikzlibrary{shapes,patterns,arrows,graphs,decorations.pathreplacing,calc}

\usepackage{xcolor,graphicx,color}
\usepackage{xspace}
\usepackage{array}
\usepackage{cite}

\usepackage[textsize=tiny,textwidth=1.25cm,color=green]{todonotes}

\newcommand{\opt}{\textsc{Opt}\xspace}

\newcommand{\eps}{\varepsilon}
\newcommand{\bigO}{\mathcal{O}}

\newcommand{\newjob}{\ensuremath{{j^{\star}}}}

\newcommand{\factor}{\ensuremath{\left(1+\frac\eps2\right)}}
\newcommand{\pj}{\ensuremath{p_{ij}}} 

\newcommand{\intx}{\ensuremath{\left[r_x, d_x - \factor p_x \right)}}

\newcommand{\assign}{\ensuremath{\leftarrow}}

\newcommand{\intervals}{\ensuremath{\mathcal{I}}}

\newcommand{\scriptS}{\ensuremath{\mathcal S}}

\title{$\boldmath{O(\sfrac1\eps)}$ is the answer in online weighted throughput maximization} 

\author{Franziska Eberle}{Technische Universität Berlin}{f.eberle@tu-berlin.de}{https://orcid.org/0000-0001-8636-9711}{Supported by the Dutch Research Council (NWO), Netherlands Vidi grant 016.Vidi.189.087.}

\authorrunning{F. Eberle} 

\Copyright{Franziska Eberle} 



\relatedversion{} 

\nolinenumbers 

\EventEditors{John Q. Open and Joan R. Access}
\EventNoEds{2}
\EventLongTitle{42nd Conference on Very Important Topics (CVIT 2016)}
\EventShortTitle{CVIT 2016}
\EventAcronym{CVIT}
\EventYear{2016}
\EventDate{December 24--27, 2016}
\EventLocation{Little Whinging, United Kingdom}
\EventLogo{}
\SeriesVolume{42}
\ArticleNo{23}

\usepackage[ruled,vlined]{algorithm2e}

\begin{document}

\maketitle

\begin{abstract}
	We study a fundamental online scheduling problem where jobs with processing times, weights, and deadlines arrive online over time at their release dates. The task is to preemptively schedule these jobs on a single or multiple (possibly unrelated) machines with the objective to maximize the weighted throughput, the total weight of jobs that complete before their deadline. To overcome known lower bounds for the competitive analysis, we assume that each job arrives with some slack~$\eps > 0$; that is, the time window for processing job~$j$ on any machine~$i$ on which it can be executed has length at least~$(1+\eps)$ times $j$'s processing time on machine~$i$. 
	
	Our contribution is a best possible online algorithm for weighted throughput maximization on unrelated machines: Our algorithm is~$\bigO\big(\frac1\eps\big)$-competitive, which matches the lower bound for unweighted throughput maximization on a single machine. Even for a single machine, it was not known whether the problem with weighted jobs is ``harder'' than the problem with unweighted jobs. Thus, we answer this question and close weighted throughput maximization on a single machine with a best possible competitive ratio~$\Theta\big(\frac1\eps\big)$. 
	
	While we focus on non-migratory schedules, our algorithm achieves the same (up to constants) performance guarantee when compared to an optimal migratory schedule. 
\end{abstract}

\newpage
\setcounter{page}{1}

\section{Introduction}
%

We consider an online scheduling problem with~$m$ parallel \emph{unrelated} machines.
Online over time, job~$j$ arrives at its \emph{release date}~$r_j$. Upon arrival of job~$j$, its \emph{processing time}, sometimes also referred to as \emph{size},~$p_{ij} \in \mathbb{R}_{> 0} \cup \{\infty\}$ on machine~$i \in [m] := \{1,\ldots,m\}$, its \emph{weight}~$w_j$, and its \emph{deadline}~$d_j$ are revealed to the online algorithm. The \emph{density} of~$j$ on machine~$i$ is given by~$\rho_{ij} := \frac{w_j}{p_{ij}}$. A machine~$i$ is \emph{eligible} for job~$j$ if~$p_{ij} < \infty$. If~$p_{ij} = p_j$ holds for all~$i$ and all~$j$, we call the machines \emph{identical} and omit the index. 

The processing of a job is allowed to be interrupted, we say \emph{preempted}, and resumed at any later point in time, also on a different machine; that is, we allow \emph{migration}. A job~$j$ completes if~$\sum_{i=1}^m \frac{q_{ij}}{p_{ij}} = 1$, where~$j$ receives a total of~$q_{ij}$ time units on machine~$i$. In a feasible schedule, at any point in time, every machine can process at most one job, and every job can be processed by at most one machine. The objective is to find a feasible schedule that maximizes the \emph{weighted throughput}, the total weight of jobs that meet their deadlines. 


Due to the lack of information because of the online nature of the problem, we cannot expect to find a throughput-optimal schedule for every instance, even on a single machine~\cite{DertouzosM89}. Thus, we employ standard competitive analysis to measure the performance of an online algorithm, where we compare the weighted throughput of an online algorithm to the weighted throughput of an optimal offline algorithm, that has complete knowledge about the instance in advance and uses this information to make scheduling decisions. 

It has been known for 30 years that in this general setting no deterministic algorithm can achieve a bounded competitive ratio on identical machines~\cite{BaruahKMMRRSW92,KorenS94}. In fact, even when allowing randomization on a single machine, the competitive ratio for weighted throughput maximization remains unbounded~\cite{CanettiI98}. 
All of the aforementioned lower bounds heavily rely on ``tight'' jobs, that is, on jobs that need to be processed immediately and continuously upon release in order to finish before their deadlines. 

To overcome these lower bounds, we make a standard \emph{slackness} assumption that the window for processing job~$j$ has some slack: we assume that $d_j - r_j \geq (1+\eps)p_{ij}$ holds if machine~$i$ is eligible for~$j$, i.e., if~$p_{ij} < \infty$. We expect our algorithm to perform better with larger slack parameter~$\eps >0$. This trade-off between slack and competitive ratio has successfully been studied before~\cite{LucierMNY13,AzarKLMNY15,SchwiegelshohnS16,GarayNYZ02,Goldwasser1999,BaruahH97,ChenEMSS2020,EberleMS23}. 

Recently, unweighted throughput maximization has been solved on a single machine with a competitive ratio of~$\bigO\big(\frac1\eps\big)$~\cite{ChenEMSS2020}, on identical machines with a~$\bigO(1)$-competitive algorithm~\cite{MoseleyPSZ22}, and on unrelated machines with a competitive ratio of~$\bigO\big(\frac1\eps\big)$~\cite{EberleMS23}. 
For weighted throughput maximization it is known that~$\bigO(1)$-competitive algorithms are not possible, independent of the machine environment, even when allowing randomization~\cite{KorenS94}. Even on a single machine, there remained a gap between the performance bound~$\bigO\big(\frac1{\eps^2}\big)$ of the algorithm by~\cite{LucierMNY13} and the lower bound~$\Omega\big(\frac1\eps\big)$ carried over from the unweighted setting~\cite{ChenEMSS2020}. 

In this work, we close this gap and give an (up to constant factors) best possible online algorithm for weighted throughput maximization on unrelated machines with competitive ratio~$\bigO\big(\frac1\eps\big)$. Our non-migratory algorithm remains~$\bigO\big(\frac1\eps\big)$-competitive against the optimal schedule that is allowed to use migration. 
In particular, we solve the problem on a single machine, matching the known lower bound for unweighted throughput maximization~\cite{ChenEMSS2020}. 


\subsection*{Related work}

Online throughput maximization gained a lot of interest during the last years~\cite{ChenEMSS2020,EberleMS23,JamalabadiSS20,MoseleyPSZ22}, but research has been active for decades~\cite{BaruahKMRRS91,BaruahHS94,BaruahKMMRRSW92,BaruahH97}.

For tight jobs with~$d_j - r_j = p_j$, there are non-constant lower bounds on the competitive ratios of deterministic and randomized algorithms~\cite{BaruahKMMRRSW92,CanettiI98}, which justify our slackness assumption. Baruah et al.~\cite{BaruahKMMRRSW92} give a lower bound of $(1 + \sqrt \mu)^2$ on the competitive ratio of any deterministic single-machine algorithm, where~$\mu = \frac{\max_j \rho_j}{\min_j \rho_j}$, while Koren and Shasha~\cite{KorenS95} give an algorithm with matching competitive ratio. On identical machines, their algorithm achieves the best possible competitive ratio of~$\Theta(\ln \mu)$~\cite{KorenS94}. Canetti and Irani~\cite{CanettiI98} consider randomized algorithms and show a lower bound of~$\Theta\big(\frac{\log \nu}{\log \log \nu}\big)$, where~$\nu = \min\big\{\frac{\max p_j}{\min p_j}, \frac{\max w_j}{\min w_j} \big\}$. They also give an almost matching upper bound. 

Since both parameters,~$\mu$ and~$\nu$, can become arbitrarily large, research started to investigate instances satisfying a slackness assumption~\cite{LucierMNY13}. Most relevant to our work is the $\bigO\big(\frac1{\eps^2}\big)$-competitive algorithm by
Lucier et al.~\cite{LucierMNY13} for weighted throughput maximization on identical machines under the slackness assumption. Azar et al.~\cite{AzarKLMNY15} study the same problem and give a truthful mechanism with a competitive ratio of~$2 + \Theta\Big(\frac1{\sqrt[3]{1+\eps}-1}  + \frac1{(\sqrt[3]{1+\eps}-1)^3} \Big)$. 

The special case of maximizing machine utilization, where the weight of each job equals its processing time, allows for~$\bigO(1)$-competitive algorithms, even in settings without slack.
On a single machine, the algorithm by Baruah et al.~\cite{BaruahKMRRS91,BaruahKMMRRSW92} is $4$-competitive, and on identical machines, Koren and Shasha~\cite{KorenS94} claim an~$\bigO(1)$-competitive algorithm. 

In the \emph{unweighted} setting, Baruah et al.~\cite{BaruahHS94} show that non-trivial competitive ratios are impossible in the presence of tight jobs. However, randomization allows for a competitive ratio of~$\bigO(1)$~\cite{KalyanasundaramP03}. 
If every job arrives with a slack of~$\eps$, the (deterministic) algorithm by Chen et al.~\cite{ChenEMSS2020} achieves the provably best competitive ratio of~$\Theta\big(\frac1\eps\big)$ on a single machine. On at least two machines, the algorithm by Moseley et al.~\cite{MoseleyPSZ22} is~$\bigO(1)$-competitive, even for instances without slack. Eberle et al.~\cite{EberleMS23} design a~$\Theta\big(\frac1\eps\big)$-competitive algorithm for throughput maximization on unrelated machines when each job has~$\eps$-slack. 

\subsection*{Our results}

As our main result, we present an~$\bigO\big(\frac1\eps\big)$-competitive algorithm for online weighted throughput maximization. 
\begin{theorem}\label{theo:mainResult}
	For weighted throughput maximization on unrelated machines without migration, there is an~$\bigO\big(\frac1\eps\big)$-competitive online algorithm. 
\end{theorem}

This generalizes and improves the~$\bigO\big(\frac1{\eps^2}\big)$-competitive algorithm by~\cite{LucierMNY13}. It matches the known lower bound of~$\Omega\big(\frac1\eps\big)$ on a single machine~\cite{ChenEMSS2020} and, thus, closes the gap that remained. 


During the analysis, we focus on comparing the non-migratory schedule obtained by our algorithm to an optimal, non-migratory schedule. Because of a recent result by Karakostas and Kolliopoulos \cite{KarakostasK23}, we know that the throughput achievable without migration is within a constant multiplicative factor of the throughput achievable using migration. Thus, our result also holds in the migratory setting. 


\begin{theorem}
	For weighted throughput maximization on unrelated machines with migration, there is an~$\bigO\big(\frac1\eps\big)$-competitive online algorithm. 
\end{theorem}

\subsection*{One threshold cannot beat $\boldmath{\Theta(\sfrac1{\eps^2})}$}

Previous results for throughput maximization use a threshold-based policy to decide about the admission of newly released and the preemption of currently running jobs~\cite{ChenEMSS2020,EberleMS23,LucierMNY13,AzarKLMNY15}. Crucially, these algorithms rely on a \emph{single} density threshold~$\gamma \in \Theta(\eps)$ to determine if a currently running job is preempted in favor of a newly released job with higher density. 

The following two examples give an intuition why a single-threshold algorithm cannot break the~$\bigO\big(\frac1{\eps^2}\big)$-barrier. 
Let~$\eps <1$ and suppose that~$\gamma \in (0,1]$ is the threshold which an algorithm uses to discard currently running jobs in favor of newly released jobs with density higher by a factor at least~$\frac1\gamma$. In both examples,~$\delta \ll 1$ is a small constant, for which we will eventually consider the limit~$\delta \rightarrow 0$, and the jobs are \emph{tight}, i.e., they satisfy~$r_j + (1+\eps)p_j = d_j$. 

\begin{example}\label{ex:smallerjobs}
	There is a single machine and~$n+1$ tight jobs with the parameters~$r_0 = 0$, $p_0 = w_0 = \rho_0 = 1$ and~$r_j = r_{j-1} + (1 - \delta) p_{j-1}$, $p_j = (\eps + \delta) p_{j-1}$ and~$\rho_j = \frac{1+\delta}{\gamma} \rho_j$ for~$j \in [n]:= \{1, \ldots, n\}$. 
	
	By our choice of parameters, job~$j$ interrupts the execution of job~$j-1$ immediately upon arrival. It is easy to calculate that~$d_j \geq d_{j-1}$ and~$C_j = d_{j-1}$ for~$j \in [n]$ if the processing of job~$j$ is not interrupted. Combining these two observations implies that the algorithm can only complete the last job on time if the constant~$\delta$ is chosen sufficiently small. Hence, the algorithm obtains a total weight of~$\rho_n p_n = \Big(\frac{(1+\delta)(\eps + \delta)}{\gamma}\Big)^n \rightarrow \Big(\frac\eps\gamma\Big)^n$ as~$\delta \rightarrow 0$. Conversely, by scheduling only job~$1$, one can obtain a total weight of~$1$, implying that the competitive ratio is at least~$\big(\frac\gamma\eps\big)^n$. Hence,~$\gamma \leq \eps$ is necessary to achieve a bounded competitive ratio. 
\end{example} 	

\begin{example}\label{ex:largerjob}
	We have a single-machine instance with two tight jobs~$1$ and~$2$. The parameters are $p_1 = w_1 =2$,~$r_1 = 0$ and~$p_2 = \frac1\eps$, $w_2 = \frac1{\eps \gamma - \delta}$,~$r_2 = \delta$. Since both jobs are tight, no feasible schedule can complete both jobs on time. Hence, it is optimal to schedule only the second job upon release and obtain a total weight of~$\frac1{\eps \gamma - \delta}$. However, the parameters are chosen such that an algorithm with threshold~$\gamma$ admits the first job upon release and cannot discard it in favor of the second job, implying a competitive ratio of~$\Omega\big( \frac1{\eps \gamma - \delta} \big)$, which goes to~$\Omega\big( \frac1{\gamma \eps} \big)$ as~$\delta \rightarrow 0$. 
\end{example}

What becomes apparent in the examples is that by relying on a single threshold to guide the admission decisions, an algorithm is both too careless (\Cref{ex:smallerjobs}) and too conservative (\Cref{ex:largerjob}) in admitting jobs. In fact, such an algorithm does not distinguish the reasons for a job having a relatively high density: it might be caused by a large weight or by a small processing time. 

We show that, by using \emph{two} distinct thresholds, a simple greedy algorithm achieves a competitive ratio of~$\Theta\big(\frac1\eps\big)$, which is optimal up to constants even in the unweighted setting~\cite{ChenEMSS2020}. Our algorithm compares the sizes of a newly released job~$\newjob$ and a currently running job~$j$, in order to decide whether to abandon the latter in favor of the former. In \Cref{ex:smallerjobs}, we have already established that if~$j$ is preempted in favor of~$\newjob$ with~$p_{i\newjob} \in \bigO(\eps)p_{ij}$, then the density of~$\newjob$ should be greater by a factor~$\Omega\big(\frac1\eps\big)$. Conversely, to avoid the issue present in \Cref{ex:largerjob}, if the new job~$\newjob$ is larger than the currently running job~$j$, then~$\newjob$ should be admitted if it has a similar density as~$j$. For technical reasons, our algorithm employs a third admission rule that smoothly interpolates between the threshold~$\Theta(\eps)$ for jobs smaller by a factor~$\bigO(\eps)$ and a threshold~$\Theta(1)$ for larger jobs. 

In the following section, we formally describe our algorithm before analyzing its competitive ratio in the two subsequent sections.

\section{The two-threshold algorithm}

In this section, we design the two-threshold algorithm. We assume without loss of generality that~$\eps \leq 1$ as otherwise we can simply run the algorithm with~$\eps = 1$ and obtain a constant competitive ratio. 

The two-threshold algorithm starts job~$j$ on machine~$i$ for the first time only before~$d_j - \factor p_{ij}$ , we say~$j$ is \emph{admitted} to machine~$i$ at time~$a_j$. For each machine~$i$, the algorithm maintains the set of jobs that are active at time~$\tau$. A job~$j$ is~\emph{active} at time~$\tau$ on machine~$i$ if it was admitted to~$i$ before time~$\tau$ and can still complete before time~$a_j + \factor p_{ij}$, i.e., the remaining processing time of~$j$ on~$i$ is at most~$a_j +  \factor p_{ij} - \tau$.

\subparagraph*{Scheduling routine} At time~$\tau$ and on each machine~$i$, the algorithm simply processes the job~$j$ which is active for~$i$ at~$\tau$ and has the highest density~$\rho_{ij} = \frac{w_j}{p_{ij}}$ among all such jobs.

\subparagraph*{Admission routine} There are two events that trigger the admission routine at time~$\tau$: the release of a new job and the completion of a job. The admission routine loops over the machines and decides whether the currently running job~$j$ should be preempted for a job with higher density. 

To this end, it considers the jobs that have been released, have not yet been admitted, and  whose deadline is sufficiently far in the future, i.e.,~$d_{\newjob} - \tau \geq \factor p_{i\newjob}$, in decreasing order of machine-dependent density~$\rho_{i\newjob}$. Let~$\newjob$ be the job with highest density that has not been considered before. The algorithm compares~$\newjob$'s processing time with that of the job~$j$ that is currently processed by machine~$i$.


If no such job exists, then~$\newjob$ is admitted to machine~$i$ and starts executing immediately. If such a job~$j$ exists and its processing time is at least~$\frac2\eps p_{i\newjob}$, then the first density-threshold~$\frac8\eps$ is invoked: if~$\rho_{i\newjob} \geq \frac8\eps \rho_{ij}$, then~$\newjob$ is admitted to~$i$ at time~$a_{\newjob} = \tau$. If~$j$ exists and~$\frac\eps2 p_{ij} < p_{i\newjob} \leq p_{ij}$, then we use a smooth transition between the two thresholds: if~$w_{\newjob} \geq 4 w_j$, then~$\newjob$ is admitted to~$i$ at time~$a_{\newjob} = \tau$. 
 Otherwise, that is,~$j$ is currently running on machine~$i$ and its processing time is smaller than~$p_{i\newjob}$, then the second density-threshold~$4$ is invoked: if~$\rho_{i\newjob} \geq 4 \rho_{ij}$, then~$\newjob$ is admitted to~$i$ at time~$a_{\newjob} = \tau$. 

If job~$\newjob$ interrupts the execution of job~$j$, we say that~$j$ is the \emph{parent}~$\pi(\newjob)$ of~$\newjob$ and~$\newjob$ is a \emph{child} of~$j$. Note that by construction, a newly admitted job has highest density on its machine and starts processing immediately. We summarize our algorithm in \Cref{alg:Alg}.

\begin{algorithm}[t]
	\DontPrintSemicolon
	\caption{Two-threshold algorithm}
	\label{alg:Alg}	
	\textbf{Initialize:} If $\eps > 1$, then reset $\eps \leftarrow 1$.\; \smallskip	
	\textbf{Scheduling Routine:} At all times $\tau$ and on all machines $i$, run the job with highest density that is active for $i$.\; \smallskip	
	\textbf{Event:} Release of a new job at time $\tau$ \;
	\hspace{1em}Call Admission Routine\;\smallskip
	\textbf{Event:} Completion of a job at time $\tau$\;
	\hspace{1em}Call Admission Routine\;\medskip 
	\textbf{Admission Routine:} \;
	\For{$i=1$ {\normalfont to} $m$}{
		$\newjob \assign \arg\max\{\rho_{ij}  \,|\, r_j \leq \tau, d_j - \tau \geq \factor p_{ij}, j \text{ not admitted or considered}\}$ \;
		$K$ $\assign$ \{$k: k$ active on machine $i$ at time $\tau$ \} \tcp*[r]{jobs currently active for $i$}
		\uIf{$K = \emptyset$}{
			admit $\newjob$ to $i$ and $a_{\newjob} \leftarrow \tau$ \;
			$\pi(\newjob) \assign \emptyset$ \tcp*[r]{$\newjob$ does not have a parent}
		}
		\Else{
			$j \assign \arg\max\{\rho_{ik} \,|\, k \in K\}$ \tcp*[r]{currently running job} 
			\uIf{ $p_{i\newjob} \leq \frac\eps 2 p_{ij}$ {\normalfont {\bfseries and} $\rho_{i\newjob} \geq \frac8\eps \rho_{ij}$} } {
				admit $\newjob$ to $i$ and $a_{\newjob} \leftarrow \tau$ \;
				$\pi(\newjob) \leftarrow j$ \tcp*[r]{parent of $\newjob$} 
			}
			\ElseIf{$ \frac\eps2 p_{ij} < p_{i\newjob} \leq p_{ij}$ {\normalfont \bfseries and} $w_{\newjob} \geq 4 w_j$ }{
				admit $\newjob$ to $i$ and $a_{\newjob} \leftarrow \tau$ \;
				$\pi(\newjob) \leftarrow j$ \tcp*[r]{parent of $\newjob$} 
			}
			\ElseIf{$p_{i\newjob} > p_{ij}$ {\normalfont {\bfseries and} $\rho_{i\newjob} \geq 4 \rho_{ij}$}}{				
				admit $\newjob$ to $i$ and $a_{\newjob} \leftarrow \tau$ \;
				$\pi(\newjob) \leftarrow j$ \tcp*[r]{parent of $\newjob$} 
			}
		}
	}
\end{algorithm}

The following observation formalizes the ``smooth transition'' between the two density thresholds. 

\begin{observation}\label{obs:smooth}
	Consider jobs~$j$ and~$k$ with~$\frac\eps2 p_{ij} < p_{ik} \leq p_{ij}$ for some machine~$i$. If~$w_k \geq 4 w_j$, then~$\rho_{ik} = \frac{w_k}{p_{ik}} \geq \frac{4 w_j}{p_{ij}} = 4 \rho_{ij}$. If~$w_k < 4 w_j$, then~$\rho_{ik} <\frac{4w_j}{\eps/2 p_{ij}} = \frac8\eps\rho_{ij}$. Further, if~$p_{ik} > p_{ij}$ and~$\rho_{ik}\geq 4 \rho_{ij}$, then~$w_k = \rho_k p_k \geq 4 w_j$. 
\end{observation}

\subsubsection*{Roadmap of the analysis}

The analysis of the two-threshold algorithm naturally splits into two parts. In \Cref{sec:finishEnoughWeight}, we show that the highest-density rule used for scheduling active jobs guarantees that the total weight of jobs completed before their deadlines is at least half of the total weight of jobs admitted by the admission routine. In \Cref{sec:acceptEnoughWeight}, we compare the total weight of the jobs admitted by the two-threshold algorithm to the weighted throughput of an optimal solution before proving \Cref{theo:mainResult}.

\section{Weight of finished jobs vs. weight of admitted jobs}\label{sec:finishEnoughWeight}

In this section, we show that the two-threshold algorithm obtains at least half of the total weight of the jobs that were admitted. We prove the following theorem where~$J$ denotes the set of jobs admitted by our algorithm and~$F \subseteq J$ the set of jobs completed before their respective deadlines. 

\begin{theorem}\label{theo:finishedWeight} 
	Let~$J$ and~$F$ be the set of jobs admitted and finished, respectively, by the two-threshold algorithm. Then, 
	\[
		\sum_{j \in F} w_j \geq \frac12 \sum_{j \in J} w_j.
	\] 
\end{theorem}

For intuition, consider an instance that only consists of a job~$j$ and the set~$K$ of~$j$'s children. Suppose that~$j$ does not finish on time as otherwise the theorem trivially holds. Recall that~$j$ was admitted at~$a_j \leq d_{j} - \factor p_{ij}$ to machine~$i$. (Jobs that are not interrupted complete before~$a_j + \factor p_{ij} \leq d_j$.) This implies that the total processing time of jobs interrupting~$j$ is at least~$\frac\eps2 p_{ij}$. If there is at least one job~$k$ with~$p_{ik} > \frac\eps2 p_{ij}$, \Cref{obs:smooth} and the admission rule for jobs with~$\frac\eps2 p_{ij} < p_{ik} \leq p_{ij}$ imply that~$w_k \geq 4 w_j$ showing the statement. If all jobs~$k$ have processing time at most~$\frac\eps2 p_{ij}$, their densities are bounded from below by~$\frac8\eps \rho_{ij}$, and their total weight is again at least~$4w_j$. 

In the formal proof of \Cref{theo:finishedWeight}, we assume the existence of an instance that violates the statement and restrict to one that is minimal with respect to the total number of jobs. The above intuition tells us that sub-instances consisting of a job and its children cannot cause the violation. In fact, we show that we can carefully merge such sub-instances into one job without changing the fact that the complete instance violates the theorem statement, which contradicts the minimality of the original instance. 

\begin{proof}[Proof of \Cref{theo:finishedWeight}]
	Let~$U = J \setminus F$ be the set of jobs admitted by the two-threshold algorithm that were discarded, i.e., that did not complete by time~$a_j + \factor \pj$. In order to show the theorem, it suffices to prove $\sum_{j \in F} w_j \geq \sum_{j \in U} w_j$. 
	
	For the sake of contradiction, we assume that there is an instance with  $\sum_{j \in F} w_j < \sum_{j \in U} w_j$. Among all such instances, we consider an instance with the smallest number of jobs. In particular, this implies that there are no jobs in the instance that were not admitted by the algorithm and there is only one machine in the instance. We show that there is another instance with strictly fewer jobs that still satisfies the above inequality, contradicting the choice of the instance. 
	
	Without loss of generality, for all jobs~$j$, we can assume that~$r_j = a_j$ and~$d_j = r_j + (1+\eps) p_j$ holds. Indeed, since the first assumption does not change the availability of a job~$j$ at time~$a_j$ or the density, the two-threshold algorithm still admits~$j$ at time~$a_j$. Further, as the algorithm discards jobs when they cannot be completed by time~$a_j + \factor p_j < d_j$, the second assumption does not change whether a job is completed on time by the algorithm. 
	
	Observe that a job that is not interrupted completes on time. Hence, the assumption~$\sum_{j \in F} w_j < \sum_{j \in U} w_j$ implies that there are jobs whose processing is interrupted. Fix a job~$j$ that is preempted but whose children's processing is not interrupted. Let~$K$ be the set of children of~$j$, and let~$\pi = \pi(j)$ if it exists. Let~$C'_{j'}$ be the last point in time that the two-threshold algorithm works on job~$j'$, which is either the completion time of~$j'$ or the point when~$j'$ was discarded because of jobs with higher densities. Denote by~$C' := \max\{\max_{k \in K} C'_k, C_j' \}$, the last point in time when~$j$ or one of its children were processed. Observe that during~$[a_j, C')$ only~$j$ and~$j$'s children are processed. 
	
	Our goal is to create a new instance where~$j$ and its children are replaced by a new job~$j^\star$. Let~$F'$ and~$U'$ denote the finished and unfinished jobs, respectively, after the replacement. We will show that the new instance satisfies the following properties: 
	\begin{enumerate}[(i)]
		\item Job $j^\star$ is admitted at~$a_j$ and completes at time~$C'$.  
		\item $\sum_{j' \in F'} w_{j'} < \sum_{j' \in U'} w_{j'}$.
		\item There are strictly fewer jobs. 
	\end{enumerate} 
	By assumption~$j$ has at least one child. Hence, property~(iii) follows trivially from our replacement. We do not make any changes to a job~$j' \notin K \cup \{j\}$. Thus, property~(i) and the assumptions on the instance imply that our changes will not influence whether such a job~$j'$ is discarded or completed by the algorithm. 
	
	We set~$p_{j^\star} = \bar p_{j} + \sum_{k \in K} p_k$, where~$\bar p_{j} \leq p_j$ is the actual amount that the two-threshold algorithm processed~$j$ in the original instance. 
	For ensuring that~$j^\star$ is available at~$a_{j}$, we set~$r_{j^\star} = a_{j}$ and~$d_{j^\star} = r_{j^\star} + (1 + \eps) p_{j^\star}$. This choice of parameters implies that, if~$j^\star$ is admitted at time~$r_{j^\star}$, it will complete at time 
	\[
		r_{j^\star} + p_{j^\star} = a_{j} + \bar p_{j} + \sum_{k \in K} p_k = C' < d_{j^\star} 
	\] 
	since no other job is released during the interval~$[a_j, C')$ in the new instance. Thus, in order to show property~(i), it suffices to show that~$\newjob$ interrupts job~$\pi$ at~$r_\newjob$. Recall that the two-threshold algorithm compares~$p_\newjob$ with~$p_\pi$ in order to decide upon admission of~$\newjob$. There are three possibilities: $p_\newjob \leq \frac\eps2 p_\pi$, $p_\newjob \in \big(\frac\eps2 p_\pi, p_\pi\big]$, and~$p_\newjob > p_\pi$. Depending on the interval, different admission rules apply.

	For defining the weight of job~$j^\star$, we distinguish two cases based on job~${j}$.
	
	\smallskip 
	\noindent \textbf{Case I:} If~${j}$ completes on time, set~$w_{j^\star} = w_{j} + \sum_{k \in K} w_k \geq w_j$. We observe that~$\sum_{k \in K} p_k \leq \frac\eps2p_j$ as~$j$ completes on time. 
	Further, 
	\[
		\rho_{j^\star} = \frac{w_{{j}} + \sum_{k \in K} w_k}{p_{j} + \sum_{k \in K} p_k} \geq \frac{\rho_{j} p_{j} + \frac8\eps \rho_{j} \sum_{k \in K} p_k }{p_{j} + \sum_{k \in K} p_k} \geq \rho_{j}. 
	\]
	Thus, if~$p_\newjob$ and~$p_j$ belong to the same interval with respect to~$p_\pi$,~$\newjob$ is admitted upon release and property~(i) is satisfied. If they belong to different intervals, we note that
	\begin{equation}\label{eq:jCompletes}
		\factor p_j \geq p_\newjob = p_j + \sum_{k \in K} p_k \geq p_j
	\end{equation} 
	and distinguish two cases. 	
	\begin{itemize}
		\item $\bm{p_j \leq \frac \eps2 p_\pi < p_\newjob:}$ We note that \eqref{eq:jCompletes} and~$\eps \leq 1$ imply~$p_\newjob \leq p_\pi$. Thus,
		\[
		w_\newjob = \rho_\newjob p_\newjob \geq \rho_j \frac \eps2 p_\pi \geq \frac8\eps \rho_\pi \frac\eps2 p_\pi = 4 w_\pi,
		\]
		which guarantees property~(i). 
		\item $\bm{p_j \leq p_\pi < p_\newjob :}$ We have~$p_\newjob = (1 + \delta)p_j \leq (1+\delta)p_\pi$ for some~$\delta \in \big(0,\frac\eps2\big]$. Further,~$w_\newjob \geq w_j + \delta \frac 8\eps w_j \geq 4 (1 + \delta)w_\pi$. Thus, \(
		\rho_\newjob = \frac{w_\newjob}{p_\newjob} \geq \frac{4 (1+\delta) w_\pi}{(1+\delta) p_\pi} \geq 4 \rho_\pi,
		\) and property~(i) holds. 	
	\end{itemize}
	Hence, the total weight of the jobs completed by the two-threshold algorithm and the total weight of the discarded jobs does not change in all cases, which implies property~(ii).
	
	\smallskip
	\noindent \textbf{Case II:} If~$j$ does not complete on time, we set~$w_{j^\star} = \frac34\sum_{k \in K} w_k$. If some~$k^\star \in K$ satisfies~$p_{k^\star} > \frac\eps2 p_j$, then~$\sum_{k \in K} w_k \geq w_{k^\star} \geq 4 w_j$ by definition of the two-threshold algorithm. Using that~$j$ does not finish on time, we know that~$\sum_{k \in K} p_k > \frac\eps2 p_j$. Thus, if the processing times for all~$k \in K$ are bounded from above by~$\frac\eps2 p_j$, then~$\sum_{k \in K} w_k \geq \frac8\eps \rho_j \sum_{k \in K} p_k \geq 4 w_j$. 
	
	For property~(i), we start by bounding~$w_\newjob$ and~$\rho_\newjob$. Using the observation above,~$w_\newjob = \frac34 \sum_{k \in K} w_k \geq 3 w_j$. By \Cref{obs:smooth},~$k \in K$ with~$p_k \in \big(\frac\eps2 p_j, p_j\big]$ satisfy~$\rho_k \geq 4 \rho_j$. Hence, \(
		\sum_{k \in K} w_k = \sum_{k \in K} \rho_k p_k \geq 4 \rho_j \sum_{k \in K} p_k. 	
	\)
	Using again that~$\sum_{k \in K} w_k \geq 3 w_j$, we have \[
		\rho_\newjob = \frac{w_\newjob}{p_\newjob} \geq \frac{1/2 \sum_{k \in K} w_k + w_j }{\bar{p_j} + \sum_{k \in K} p_k} \geq \frac{2\rho_j\sum_{k \in K} p_k + \rho_j p_j
		}{\sum_{k \in K} p_k + \bar{p_j}} \geq \rho_j. 
	\]
	As before, if~$j$ and~$\newjob$ are in the same interval with respect to~$p_\pi$, these observations guarantee that~$\newjob$ interrupts~$\pi$ at~$a_j$, which implies property~(i). If they belong to different intervals, we distinguish five cases. 
	\begin{itemize}		
		\item $\bm{p_j \leq \frac\eps2 p_\pi < p_\newjob \leq p_\pi:}$ We have $		w_\newjob = \rho_\newjob p_\newjob \geq \rho_j \frac\eps 2 p_\pi \geq \frac8\eps \rho_\pi \frac\eps2p_\pi = 4 w_\pi$. 
		\item $\bm{p_j \leq \frac \eps2 p_\pi < p_\pi < p_\newjob:}$ We have $\rho_\newjob \geq \rho_j > 4 \rho_\pi$. 
		\item $\bm{\frac\eps2 p_\pi < p_j \leq p_\pi < p_\newjob:}$ We know that $\rho_\newjob \geq \rho_j = \frac{w_j}{p_j} \geq \frac{4 w_\pi}{p_\pi} = 4 \rho_\pi$. 
		\item $\bm{p_\newjob \leq \frac\eps2 p_\pi < p_j:}$ We note that~$p_\newjob \geq \sum_{k \in K} p_k > \frac\eps2 p_j$, which implies~$p_j \leq p_\pi$. We have~$\rho_\newjob = \frac{w_\newjob}{p_\newjob} \geq \frac{3 w_j}{\eps/2 p_\pi} \geq \frac{12w_\pi}{\eps/2 p_\pi} = \frac{24}{\eps}\rho_\pi$. 
		\item $\bm{\frac\eps2 p_\pi < p_\newjob \leq p_\pi < p_j:}$ We have~$w_\newjob \geq 3 w_j = 3 \rho_j p_j > 3 \cdot 4 \rho_\pi p_\pi = 12 w_\pi$. 		
	\end{itemize}

Hence, in all cases,~$\newjob$ satisfies the conditions of the two-threshold algorithm to interrupt the processing of~$\pi$ at~$r_\newjob$.

Recall that~$\sum_{k \in K} w_k \geq 4  w_j$. After replacing~$K \cup \{j\}$ with~$\newjob$, it holds that 
\[
	\sum_{j \in F'} w_{j} = \sum_{j \in F} w_{j} - \frac14 \sum_{k \in K} w_k < \sum_{j \in F} w_{j} - w_{j} < \sum_{j \in U} w_{j} - w_{j} = \sum_{j \in U'} w_{j},
\]
which implies property~(ii). 

As argued above, this contradicts the choice of the instance, which concludes the proof. 
\end{proof}

\section{Weight of admitted jobs vs. weight of \textsc{Opt}}\label{sec:acceptEnoughWeight}

In this section, we show that the total weight of jobs finished by an optimal solution is up to a factor~$\mathcal O \big( \frac1\eps \big)$ comparable to the total weight of jobs admitted by the two-threshold algorithm: 

\begin{theorem}\label{theo:enoughWeight}
	Let \opt and~$J$ be the set of jobs admitted by an optimal, non-migratory solution and the two-threshold algorithm, respectively. Then, 
		$\sum_{x \in \opt} w_x \leq \mathcal{O}\big( \frac1\eps \big) \sum_{j \in J} w_j$. 
\end{theorem}

For proving this statement, it is sufficient to focus on~$X$, the set of jobs scheduled by \opt that the two-threshold algorithm did not admit since~$\opt \subseteq X \cup J$. 

Fix a job~$x \in X$ that \opt schedules on machine~$i$. The two-threshold algorithm admits a job~$j$ during the interval~$[r_j, d_j - \factor p_{ij})$ if it is sufficiently dense. Since~$x$ is not admitted by our algorithm, the algorithm is processing jobs~$J_x$ on machine~$i$ during the interval~$[r_x, d_x - \factor p_{ix})$ with densities that are large and prevent interruption by~$x$.
That is, for jobs~$j \in J_x$ with~$\frac\eps2 p_{ij} \geq  p_{ix}$ it holds that~$\rho_{ij} > \frac\eps8 \rho_{ix}$, for jobs~$j \in J_x$ with~$p_{ix} \in \big(\frac\eps2 p_{ij}, p_{ij}]$ \Cref{obs:smooth} implies~$\rho_{ij} > \frac\eps8 \rho_{ix}$ and for jobs~$p_{ij} < p_{ix}$ it holds that~$\rho_{ij} > \frac14 \rho_{ix}$. We say that the jobs~$J_x$ \emph{block} the admission of~$x$. We will charge the weight~$w_x$ to the weight of the jobs in~$J_x$. Exploiting the two thresholds which the algorithm uses to make admission decisions, we show that the algorithm ``obtains'' a weight from partially finished jobs of at least~$\Omega(\eps) w_x$ in the interval~$[r_x, d_x)$.

\subparagraph*{Proof idea}
We give an intuition by considering a single-machine instance where the two-threshold algorithm admits exactly one job~$j$. Consider the jobs in~$X$ whose admission was blocked by~$j$: We know that the interval ~$[r_x, d_x - \factor p_x)$ is completely covered by the processing of job~$j$, i.e., by the interval~$[a_j, C_j)$. 

Now consider a job~$x$ with~$p_x \leq p_j$. Thus, the deadline of~$x$ is at most~$C_j + \factor p_x \leq a_j + \big(2 + \frac\eps2) p_j$. This implies that \opt can schedule such jobs only during~$\big[ a_j, a_j + \big(2 + \frac\eps2\big) p_j \big)$, an interval of length~$\big(2 + \frac\eps2\big) p_j$. The admission rule for the case~$p_x \leq \frac\eps2 p_j$ and \Cref{obs:smooth} for~$p_x \in \big(\frac\eps2 p_j, p_j]$ imply~$\rho_x < \frac 8 \eps \rho_j$. Hence, 
\[
	\sum_{\substack{x \in X \\ p_x \leq p_j }} w_x = \sum_{\substack{x \in X \\ p_x \leq p_j }} \rho_x p_x \leq \frac8\eps \rho_j \sum_{\substack{x \in X \\ p_x \leq p_j }} p_x \leq \frac8\eps \rho_j \bigg(2 + \frac\eps2\bigg) p_j = \bigg(\frac{16}\eps + 4\bigg)  w_j. 
\]

For a job~$x$ with~$p_x > p_j$, the slacknes assumption guarantees that~$r_x \leq d_x - (1+\eps) p_x$. Further, the interval~$\big[r_x, d_x - \factor p_x\big)$ is contained in~$[a_j, C_j)$. This allows us to upper bound the processing time~$p_x$ by~$\frac2\eps p_j$. Thus, \opt can schedule such jobs only during~$\big[a_j, a_j + \big(1 + \frac2\eps\big) p_j\big)$. Using the admission rule of our algorithm in this case gives \[
	\sum_{\substack{x \in X \\ p_x > p_j}} w_x = \sum_{\substack{x \in X \\ p_x > p_j}} \rho_x p_x < 4 \rho_j \sum_{\substack{x \in X \\ p_x > p_j}} p_x < 4 \rho_j \bigg(1 + \frac2\eps\bigg) p_j = \bigg(4 + \frac8\eps \bigg) w_j. 
\]
Combining the above two calculations yields that~$\sum_{x \in X} w_x \in \mathcal O\big(\frac1\eps w_j\big) $. 

\subparagraph*{Proof outline}
In this particular instance, each job~$x \in X$ is blocked by a job with either larger or smaller processing time. In general, this is not necessarily true. Hence, in order to extend this idea to arbitrary instances, we partition the jobs in~$X$ according to whether at least half of their \emph{availability interval}~$\intx$ is covered by jobs in~$J$ with smaller or larger processing times. We then show that by losing an additional factor~$3$, we can assume that only one type covers the availability interval of each job in~$X$. This is done in \Cref{lem:coverAvailabilityInterval}. 

An additional technical challenge poses the fact that, even after we assume that a job is blocked by either shorter or longer jobs, the densities of these jobs are still not uniform enough to directly generalize the above idea to arbitrary instances. In \Cref{lem:DensityClasses}, we show that, at the loss of an additional factor~$4$, we can partition the jobs in~$X$ according to their densities and bound the weight for each density level separately. We then upper bound \opt's available time for scheduling jobs of a certain density level in \Cref{lem:AvailableTime:SmallJobs,lem:AvailableTime:BigJobs} depending on the size of the blocking jobs.

\subparagraph*{Proof of \Cref{theo:enoughWeight}}
Since \opt and the two-threshold algorithm are non-migratory, we fix one particular machine~$i$ and only consider jobs that either \opt or the two-threshold algorithm scheduled on machine~$i$. For simplicity, we drop the index~$i$ for the remainder of this section.

We first argue that we can focus on one density level~$\ell$ at the loss of a constant factor. 

\begin{lemma}\label{lem:DensityClasses}
	Suppose there is a scheme that charges job~$j \in J$ a weight of at most~$2^\ell c p_j$ at level~$\ell \leq \lceil \log_2 \rho_j \rceil$ and no weight at level~$\ell > \lceil \log_2 \rho_j \rceil$, where~$c >0$ is a constant. Then, the total weight charged to~$j$ is at most~$4c w_j$. 
\end{lemma}

\begin{proof}
	The total weight charged to~$j$ is upper bounded by \[
	\sum_{-\infty}^{\lceil \log_2 \rho_j \rceil} 2^\ell c p_j = c \cdot p_j \Big(1 +  2^{\lceil \log_2 \rho_j \rceil + 1} -1\Big) \leq 4 c \rho_j p_j = 4c w_j.  
	\]
\end{proof}

Having this lemma at hand, we now restrict to one density level~$\ell \in \mathbb Z$ and define~$J_{\ell} := \{ j \in J: \rho_j \geq 2^\ell \}$. 
Next, we remove the technical challenge that a job~$x \in X$ can be blocked by jobs with smaller and larger processing times. 
To this end, we carefully modify the intervals where jobs in~$J_\ell$ are scheduled such that the availability interval of a job~$x\in X$ blocked by jobs in~$J_{\ell}$ is completely contained in the modified intervals. To this end, we fix a level~$\ell$ and let~$\mathcal S_{\ell}$ denote the set of processing intervals of the jobs in~$J_\ell$, that is, the intervals during which jobs in~$J_\ell$ are processed. 

The modification works as follows: We copy each interval in~$I \in \mathcal S_\ell$ twice and call one copy the \emph{early} and the other the \emph{late} copy. For each original interval~$I= [\alpha, \omega)$, we move the early copy earlier such that it ends at~$\alpha$ and we move the late copy later such that it begins at~$\omega$. If a copy intersects with another original (even if only partially), by potentially splitting the copy, we shift the part that intersects further into the indicated direction; that is, for the early copy, we move the part earlier and for the late copy, we move the part later. We treat the time points where multiple copies overlap similarly. More precisely, if the interval~$[t,t')$ is currently contained in~$k$ different copies, we use a~$\frac1k$-fraction from every copy to cover~$[t,t')$ and send the remaining~$\frac{k-1}k$-fraction into the directions indicated by their name.

We denote the resulting {set} of intervals (including the original ones) by~$\mathcal I_\ell$. Next, we prove some structural properties about~$\mathcal I_\ell$ and, for each job~$x \in X$, relate its availability interval $\big[r_x, d_x - \factor p_x \big)$ to~$\mathcal I_{\ell_x}$ for some carefully chosen~$\ell_x \in \mathbb Z$.

\begin{observation}\label{obs:coverage}
	Let~$\scriptS$ and~$\mathcal T$ be two sets of intervals such that for each~$S \in \scriptS$ there is a~$T \in \mathcal T$ with~$S \subseteq T$. Then, the result of the modification of~$ \mathcal T$ covers all time points covered by the result of the modification of~$\scriptS$.  
\end{observation}

\begin{lemma}\label{lem:coverAvailabilityInterval}
	For each job~$x \in X$, let~$\ell_{1x} = \frac14  \cdot 2^{\lfloor \log_2 \rho_x \rfloor }$ and~$\ell_{2x} = \frac\eps8 \cdot 2^{\lfloor \log_2 \rho_x \rfloor }$. Then,~$\big[r_x, d_x - \factor p_{x}\big) \subseteq \bigcup_{I \in \intervals_{\ell_{1x}}} I$ or~$\big[r_x, d_x - \factor p_{x}\big) \subseteq \bigcup_{I \in \intervals_{\ell_{2x}}} I$. 
\end{lemma}

\begin{proof}
	Fix a job~$x \in X$ and the two levels~$\ell_{1x}$ and~$\ell_{2x}$ from the lemma statement. By assumption,~$x$ is blocked by jobs in~$J$ at all times in~$\big[r_x, d_x - \factor p_x \big)$. 
	
	Assume that~$x$ is blocked for at least half of the time by jobs~$j$ with~$p_j <  p_x$. By definition of \Cref{alg:Alg}, this implies that~$4 \rho_j > \rho_x $ holds for these jobs~$j$. Hence,~$j \in J_{\ell_{1x}}$. We will show that in this case~$\ell_{1x}$ satisfies the lemma; for simplicity, set~$\ell = \ell_{1x}$. 
	
	Conversely, suppose that~$x$ is blocked for at least half of the time by jobs~$j$ with~$p_x \leq p_j$. \Cref{obs:smooth} for~$\frac\eps2 p_j < p_x \leq p_j$ and the admission threshold for~$\frac\eps2 p_j \geq p_x$ guarantee~$\rho_x < \frac8\eps \rho_j$.  Hence,~$j \in J_{\ell_{2x}}$ holds if~$p_x \leq p_j$ and~$j$ blocks~$x$. 
	For simplicity, set~$\ell = \ell_{2x}$ in this case. 
	
	Using \Cref{obs:coverage}, it suffices to focus on the set~$\scriptS$ of intervals that actually cover the interval~$\intx$ and correspond to scheduling times of jobs that block~$x$ at level~$\ell$. By truncating, we assume that the earliest interval in~$\scriptS$ starts not earlier than~$r_x$ and that the latest interval ends no later than~$d_x - \factor p_x$. We index the intervals in~$\scriptS$ by starting point and let~$K = |\scriptS|$. Denote by~$\alpha_k$ and~$\omega_k$ the start and end point, respectively, of the~$k$th interval.
	
	By assumption,~$\scriptS$ covers at least half of~$\intx$. Thus, 
	\[
		\omega_K - \alpha_1 \leq d_x - \factor p_x - r_x \leq 2 \sum_{k=1}^K (\omega_k - \alpha_k). 
	\]
	This implies that~$\scriptS$ and its copies cover the intervals~$[\alpha_1, \alpha_1 + 2 \sum_{k = 1}^K (\omega_k - \alpha_k) )$ and~$[\omega_K - 2 \sum_{k=1}^K (\omega_k - \alpha_k), \omega_K)$ because~$\scriptS$ and the late copies would suffice to cover the former and~$\scriptS$ and the early copies would suffice to cover the latter interval. 
	
	Hence, the lemma follows if we show that~$r_x \geq \omega_K - 2 \sum_{k=1}^K (\omega_k - \alpha_k)$ and~$d_x - \factor p_x \leq \alpha_1 + 2 \sum_{k = 1}^K (\omega_k - \alpha_k)$. To this end, we observe that 
	\begin{align*}
		\left(\alpha_1 - r_x \right) + \left( \left( d_x - \factor p_x \right) - \omega_K \right)
		& = \left( \left(d_x - \factor p_x \right) - r_x \right) - (\omega_K - \alpha_1) \\
		& \leq 2 \sum_{k=1}^K (\omega_k - \alpha_1) - (\omega_K - \alpha_1),
	\end{align*}
	where we used that $\alpha_1 \geq r_x$ and~$\omega_K \leq d_x - \factor p_x$ by assumption on~$\scriptS$. This implies that both summands on the left hand side are bounded by the term on the right hand side. Rearranging shows the above bounds on~$r_x$ and~$d_x - \factor p_x$ and proves the lemma. 
\end{proof}

\begin{lemma}\label{prop:ContiguousIntervals}
	For each job~$j \in J_\ell$,~$j$'s processing intervals are contained in one contiguous interval of~$\bigcup_{I \in \intervals_\ell} I$. 
\end{lemma}

\begin{proof}
	The statement holds trivially if~$j$ only has one processing interval as this interval is in~$\intervals_\ell$. If~$j$ is preempted at some time~$\tau$ and resumed at some later time~$\tau'$, then the two-threshold algorithm processes higher-density jobs in the interval~$[\tau,\tau')$. By definition, these higher-density jobs are in~$J_\ell$ if~$j\in J_\ell$. Hence, the processing intervals of~$j$ together with these higher-density jobs form a contiguous interval in~$\bigcup_{I \in \intervals_\ell} I$. 
\end{proof}

Consider subsets of jobs of~$J_\ell$ that are maximal in the sense that the processing intervals of the corresponding jobs and their copies form a contiguous interval in~$\bigcup_{I\in\intervals_\ell} I$. \Cref{prop:ContiguousIntervals} and \Cref{lem:coverAvailabilityInterval} imply the following corollary.

\begin{corollary}\label{cor:MaximalSubset}
	The maximal subsets of jobs of~$J_\ell$ as defined above partition~$J_\ell$. Further, each job~$x \in X$ is blocked by at most one such subset. 
\end{corollary}

We now partition~$X$ as follows: Let~$X_L \subseteq X$ and~$X_S \subseteq X$ be the jobs in \opt that are, for at least half of their availability interval \intx\, blocked by jobs with larger and smaller processing times, respectively. Let~$X_{*\ell} \subset X_*$ for~$* \in \{L,S\}$ be the jobs that are blocked at level~$\ell$. 
%
The previously proven structural properties allow us to upper bound the total time that \opt has available for processing jobs in $X_{S\ell}$ and~$X_{L\ell}$ in the following two lemmas.

\begin{lemma}\label{lem:AvailableTime:SmallJobs}
	For each level~$\ell \in \mathbb Z$, the total time that \opt has available for processing jobs in~$X_{L\ell}$ 
	is at most~$\left(4 + \frac\eps2 \right) \sum_{j \in J_\ell} p_j$. 
\end{lemma} 


\begin{proof}	
	By \Cref{cor:MaximalSubset} it suffices to separately show the lemma for each maximal subset~$J'$. 
	
	Consider a job~$x$ that is blocked by a subset of~$J'$ for at least half of~$\intx$, where the jobs in~$J'$ blocking~$x$ have a larger processing time than~$x$. By the definition of the two-threshold algorithm, this implies that there is a job~$j \in J'$ with~$p_x \leq p_j$ that is processed during~$\intx$.	
	By \Cref{lem:coverAvailabilityInterval},~$\intx \subseteq \bigcup_{I \in \intervals_\ell} I$ and, therefore, $\intx$ is contained in the interval~$I = [\alpha, \omega)$ associated with the jobs in~$J'$. Combining these two observations implies that 
	\[
		[r_x, d_x) = \intx \cup \left[d_x - \factor p_x, d_x \right) \subseteq I \cup \bigg[\omega, \omega + \factor \sum_{j \in J'} p_j\bigg). 
	\]
	Using that the length of~$I$ is at most~$3 \sum_{j \in J'} p_j$ concludes the proof. 
\end{proof}

\begin{lemma}\label{lem:AvailableTime:BigJobs}
	For each level~$\ell \in \mathbb Z$, the total time that \opt has available for processing jobs in~$X_{S\ell}$
	is at most~$\left( \frac4\eps + 5 \right)  \sum_{j \in J_\ell} p_j$. 
\end{lemma} 

\begin{proof}	
	By \Cref{cor:MaximalSubset} it suffices to separately show the lemma for each maximal subset~$J'$. Let~$I = [\alpha, \omega)$ be the interval associated with~$J'$.  
	
	Consider a job~$x$ that is blocked by a subset of~$J'$ for at least half of~$\intx$, where the jobs in~$J'$ blocking~$x$ have a smaller processing time than~$x$. By definition of the two-threshold algorithm, this implies that there is a job~$j \in J'$ that is processed during~$\intx$. 
	
	By our slackness assumption, it holds that $d_x - r_x \geq (1 + \eps) p_x$ or equivalently, $p_x \leq \frac2\eps\left( d_x - \factor p_x - r_x\right)$. Since~$x$ is blocked for at least half of~$\intx$ by jobs in~$J'$, this implies $p_x \leq \frac4\eps \sum_{j \in J'} p_j$.	
	Thus, \begin{align*}
		[r_x, d_x) & = \intx \cup \left[d_x - \factor p_x, d_x \right) \\ 
		& \subseteq I \cup 
		\bigg[\omega, \omega + \left(\frac4\eps + 2 \right) \sum_{j \in J'} p_j \bigg).
	\end{align*}
	Using again that the length of~$I$ is at most~$3\sum_{j \in J'} p_j$ concludes the proof. 
\end{proof}

\begin{proof}[Proof of \Cref{theo:enoughWeight}]
	We now bound the weight of the sets $X_{S\ell}$ and $X_{L\ell}$ separately for each~$\ell \in \mathbb Z$. 
	
	By \Cref{lem:AvailableTime:BigJobs}, the time available for processing jobs in~$X_{S\ell}$ is bounded from above by~$\left(\frac4\eps + 5 \right) \sum_{j \in J_\ell} p_j$.  Being blocked at level~$\ell$ by smaller jobs implies that~$\rho_x \leq 4 \cdot 2 \cdot 2^\ell = 8 \cdot 2^\ell$. Hence, 
	\begin{equation}
		\sum_{x \in X_{S\ell}} w_x = \sum_{x \in X_{S\ell}} \rho_x p_x \leq 8 \cdot 2^\ell \left(\frac4\eps + 5 \right) \sum_{j \in J_\ell} p_j. 
	\end{equation}

	Similarly, by \Cref{lem:AvailableTime:SmallJobs}, the time available for processing jobs in~$X_{L\ell}$ is upper bounded by~$\left(4 + \frac\eps2 \right) \sum_{j \in J_\ell} p_j$ and being blocked at level~$\ell$ by larger jobs implies~$\rho_x \leq \frac{8}\eps \cdot 2 \cdot  2^\ell$, where we used \Cref{obs:smooth} for a blocking job~$j$ with~$p_{ix} \in \big(\frac\eps2 p_{ij}, p_{ij}\big]$. Hence, 
	\begin{equation}
		\sum_{x \in X_{L\ell}} w_x = \sum_{x \in X_{L\ell}} \rho_x p_x \leq \frac{16}\eps \cdot 2^\ell \left(4 + \frac \eps2 \right) \sum_{j \in J_\ell} p_j. 
	\end{equation}
	
	Combining the last two equations with \cref{lem:DensityClasses} gives 
	\[
		\sum_{x \in X} w_x \leq 4 \left( 8 \cdot \left(\frac4\eps + 5 \right)  + \frac{16}\eps \left(4 + \frac \eps2\right)\right) \sum_{j \in J} w_j = \mathcal O \left( \frac1\eps\right) \sum_{j \in J} w_j. 
	\]
\end{proof}

\subsection*{Proof of main result}

\begin{proof}[Proof of \Cref{theo:mainResult}]
	Recall that \opt is the set of jobs scheduled in an optimal non-migratory solution and that~$F$ is the set of jobs that the two-threshold algorithm completes on time. Combining \Cref{theo:finishedWeight,theo:enoughWeight}, we obtain 
	\[
		\sum_{x \in \opt} w_x \leq \sum_{x \in X} w_x + \sum_{x \in J} w_x \leq \bigO \left(\frac1\eps\right) \sum_{j \in J} w_j \leq \bigO \left(\frac1\eps \right) \sum_{j \in F} w_j.
	\]
\end{proof}

\section{Conclusion}

We have presented a provably best possible non-migratory algorithm for online weighted throughput maximization on unrelated machines, that is~$\bigO\big(\frac1\eps\big)$-competitive even against an optimal migratory scheduling using a result by~\cite{KarakostasK23}. 
Even for a single machine, only an~$\bigO\big(\frac1{\eps^2}\big)$-competitive algorithm was previously known~\cite{LucierMNY13} while the lower bound was~$\Omega\big(\frac1\eps\big)$~\cite{ChenEMSS2020}. Our result closes this gap on a single machine. 


In contrast to special cases such as maximizing throughput with unit weights~\cite{MoseleyPSZ22} or maximizing machine utilization ($w_j = p_j$)~\cite{KorenS94}, it is known that~$\bigO(1)$-competitive algorithms are not possible even on identical machines and even when using randomization~\cite{CanettiI98}. It is conceivable that~$o\big(\frac1\eps\big)$-competitive algorithms are possible for~$m \geq 2$ identical machines, which we leave as an interesting open problem. 

\bibliography{throughput}

\begin{thebibliography}{10}

\bibitem{AzarKLMNY15}
Yossi Azar, Inna Kalp{-}Shaltiel, Brendan Lucier, Ishai Menache, Joseph Naor,
  and Jonathan Yaniv.
\newblock Truthful online scheduling with commitments.
\newblock In {\em {EC}}, pages 715--732. {ACM}, 2015.
\newblock \href {https://doi.org/10.1145/2764468.2764535}
  {\path{doi:10.1145/2764468.2764535}}.

\bibitem{BaruahH97}
Sanjoy~K. Baruah and Jayant~R. Haritsa.
\newblock Scheduling for overload in real-time systems.
\newblock {\em {IEEE} Trans. Computers}, 46(9):1034--1039, 1997.
\newblock \href {https://doi.org/10.1109/12.620484}
  {\path{doi:10.1109/12.620484}}.

\bibitem{BaruahHS94}
Sanjoy~K. Baruah, Jayant~R. Haritsa, and Nitin Sharma.
\newblock On-line scheduling to maximize task completions.
\newblock In {\em {RTSS}}, pages 228--236. {IEEE} Computer Society, 1994.
\newblock \href {https://doi.org/10.1109/REAL.1994.342713}
  {\path{doi:10.1109/REAL.1994.342713}}.

\bibitem{BaruahKMMRRSW92}
Sanjoy~K. Baruah, Gilad Koren, Decao Mao, Bhubaneswar Mishra, Arvind
  Raghunathan, Louis~E. Rosier, Dennis~E. Shasha, and Fuxing Wang.
\newblock On the competitiveness of on-line real-time task scheduling.
\newblock {\em Real-Time Systems}, 4(2):125--144, 1992.
\newblock \href {https://doi.org/10.1007/BF00365406}
  {\path{doi:10.1007/BF00365406}}.

\bibitem{BaruahKMRRS91}
Sanjoy~K. Baruah, Gilad Koren, Bhubaneswar Mishra, Arvind Raghunathan, Louis~E.
  Rosier, and Dennis~E. Shasha.
\newblock On-line scheduling in the presence of overload.
\newblock In {\em {FOCS}}, pages 100--110. {IEEE} Computer Society, 1991.
\newblock \href {https://doi.org/10.1109/SFCS.1991.185354}
  {\path{doi:10.1109/SFCS.1991.185354}}.

\bibitem{CanettiI98}
Ran Canetti and Sandy Irani.
\newblock Bounding the power of preemption in randomized scheduling.
\newblock {\em {SIAM} J. Comput.}, 27(4):993--1015, 1998.
\newblock \href {https://doi.org/10.1137/S0097539795283292}
  {\path{doi:10.1137/S0097539795283292}}.

\bibitem{ChenEMSS2020}
Lin Chen, Franziska Eberle, Nicole Megow, Kevin Schewior, and Cliff Stein.
\newblock A general framework for handling commitment in online throughput
  maximization.
\newblock {\em Math. Prog.}, 183:215--247, 2020.
\newblock \href {https://doi.org/10.1007/s10107-020-01469-2}
  {\path{doi:10.1007/s10107-020-01469-2}}.

\bibitem{DertouzosM89}
Michael~L. Dertouzos and Aloysius~K. Mok.
\newblock Multiprocessor on-line scheduling of hard-real-time tasks.
\newblock {\em {IEEE} Trans. Software Eng.}, 15(12):1497--1506, 1989.
\newblock \href {https://doi.org/10.1109/32.58762}
  {\path{doi:10.1109/32.58762}}.

\bibitem{EberleMS23}
Franziska Eberle, Nicole Megow, and Kevin Schewior.
\newblock Online throughput maximization on unrelated machines: Commitment is
  no burden.
\newblock {\em ACM Trans. Algorithms}, 19(1), feb 2023.
\newblock \href {https://doi.org/10.1145/3569582} {\path{doi:10.1145/3569582}}.

\bibitem{GarayNYZ02}
Juan~A. Garay, Joseph Naor, B{\"{u}}lent Yener, and Peng Zhao.
\newblock On-line admission control and packet scheduling with interleaving.
\newblock In {\em {INFOCOM}}, pages 94--103. {IEEE} Computer Society, 2002.
\newblock \href {https://doi.org/10.1109/INFCOM.2002.1019250}
  {\path{doi:10.1109/INFCOM.2002.1019250}}.

\bibitem{Goldwasser1999}
Michael~H. Goldwasser.
\newblock Patience is a virtue: The effect of slack on competitiveness for
  admission control.
\newblock {\em J. Sched.}, 6(2):183--211, 2003.
\newblock \href {https://doi.org/10.1023/A:1022994010777}
  {\path{doi:10.1023/A:1022994010777}}.

\bibitem{JamalabadiSS20}
Samin Jamalabadi, Chris Schwiegelshohn, and Uwe Schwiegelshohn.
\newblock Commitment and slack for online load maximization.
\newblock In {\em {SPAA}}, pages 339--348. {ACM}, 2020.
\newblock \href {https://doi.org/10.1145/3350755.3400271}
  {\path{doi:10.1145/3350755.3400271}}.

\bibitem{KalyanasundaramP03}
Bala Kalyanasundaram and Kirk Pruhs.
\newblock Maximizing job completions online.
\newblock {\em J. Algorithms}, 49(1):63--85, 2003.
\newblock \href {https://doi.org/10.1016/S0196-6774(03)00074-9}
  {\path{doi:10.1016/S0196-6774(03)00074-9}}.

\bibitem{KarakostasK23}
George Karakostas and Stavros~G. Kolliopoulos.
\newblock Approximation algorithms for maximum weighted throughput on unrelated
  machines.
\newblock In {\em {APPROX/RANDOM}}, volume 275 of {\em LIPIcs}, pages
  5:1--5:17. Schloss Dagstuhl - Leibniz-Zentrum f{\"{u}}r Informatik, 2023.

\bibitem{KorenS94}
Gilad Koren and Dennis~E. Shasha.
\newblock {MOCA:} {A} multiprocessor on-line competitive algorithm for
  real-time system scheduling.
\newblock {\em Theor. Comput. Sci.}, 128(1\&2):75--97, 1994.
\newblock \href {https://doi.org/10.1016/0304-3975(94)90165-1}
  {\path{doi:10.1016/0304-3975(94)90165-1}}.

\bibitem{KorenS95}
Gilad Koren and Dennis~E. Shasha.
\newblock D\textsuperscript{over}: An optimal on-line scheduling algorithm for
  overloaded uniprocessor real-time systems.
\newblock {\em {SIAM} J. Comput.}, 24(2):318--339, 1995.
\newblock \href {https://doi.org/10.1137/S0097539792236882}
  {\path{doi:10.1137/S0097539792236882}}.

\bibitem{LucierMNY13}
Brendan Lucier, Ishai Menache, Joseph Naor, and Jonathan Yaniv.
\newblock Efficient online scheduling for deadline-sensitive jobs: Extended
  abstract.
\newblock In {\em {SPAA}}, pages 305--314. {ACM}, 2013.
\newblock \href {https://doi.org/10.1145/2486159.2486187}
  {\path{doi:10.1145/2486159.2486187}}.

\bibitem{MoseleyPSZ22}
Benjamin Moseley, Kirk Pruhs, Clifford Stein, and Rudy Zhou.
\newblock A competitive algorithm for throughput maximization on identical
  machines.
\newblock In {\em {IPCO}}, volume 13265 of {\em Lecture Notes in Computer
  Science}, pages 402--414. Springer, 2022.

\bibitem{SchwiegelshohnS16}
Chris Schwiegelshohn and Uwe Schwiegelshohn.
\newblock The power of migration for online slack scheduling.
\newblock In {\em {ESA}}, volume~57 of {\em LIPIcs}, pages 75:1--75:17. Schloss
  Dagstuhl - Leibniz-Zentrum f{\"{u}}r Informatik, 2016.
\newblock \href {https://doi.org/10.4230/LIPIcs.ESA.2016.75}
  {\path{doi:10.4230/LIPIcs.ESA.2016.75}}.

\end{thebibliography}
\end{document}